 \documentstyle[epsf,fleqn,12pt]{article}
 \renewcommand{\d}{{\mathrm d}}

\begin{document}

 %----------------------- title page --------------------------

 \thispagestyle{empty}
 \vspace*{-1 cm}

 \vspace*{-1.2in}
 \begin{flushright}
 {SHEP 96-06} \\
 February 1996 \\
 \end{flushright}

 \vspace*{1.0 in}

 \begin{center}
 {\Large \bf How to Run the Coupling in the Dipole Approach 
             to the BFKL Equation } \\
 \vspace*{1cm} {\bf K.D.  Anderson} \\
 \vspace*{.3cm}
 {\bf D.A. Ross} \\ \vspace*{.3cm}
  and \\
 \vspace*{0.3cm}
 {\bf M.G. Sotiropoulos}\\ \vspace*{.6cm}
 {\it Physics Department, University of Southampton} \\
 {\it Highfield, Southampton SO17 1BJ, U.K.} \\
 \vspace*{3 cm}
 {\bf ABSTRACT} \\ 
 \end{center}
 We use the dipole expansion to provide a systematic way of including
 the running coupling into the BFKL equation. In terms of a Borel
 representation, we obtain an expression for the kernel of the BFKL
 equation. 

 \noindent
 \underline{\ \ \ \ \ \ \ \  }
 \begin{small}
 \begin{tabbing}
 E-mail: \=mgs\=@hep.phys.soton.ac.uk \kill\\
 E-mail:  \> kda\>@hep.phys.soton.ac.uk \\
          \> dar\>@hep.phys.soton.ac.uk \\
          \> mgs\>@hep.phys.soton.ac.uk 
 \end{tabbing}
 \end{small}

 %-------------- main text -------------------------------------
 
 \parskip 0.3cm
 \vspace*{3cm}
 \vfill\eject

 \setcounter{page}{1}

 \voffset -1in
 \vskip2.0cm
 
 \noindent
 It was first pointed out by Lipatov \cite{lip86} that the running of the 
 coupling plays an  important role in the BFKL equation \cite{bfkl}. 
 When this is taken into account the eigenfunctions $\phi_i(\vec{k})$
 of the BFKL kernel ${\mathcal K}$, given by
 \begin{equation}
 \hspace{-0.95cm}
 \int \d^2 \vec{ k'} \, \alpha_s \, {\mathcal K}(\vec{k},\vec{k'}) \,
 \phi_i(\vec{k'}) = \lambda_i \, \phi_i(\vec{k\,}),  
 \label{eq1} 
 \end{equation}
 cease to oscillate beyond a certain critical value of the transverse
 momentum $\vec{k}$ of the $t$-channel gluons in the ``ladder''. 
 Above this critical value the eigenfunctions decay exponentially 
 and this decay can be matched to the oscillations below the critical
 value. 
 This phase matching, together with a phase fixing for small values 
 of the transverse momentum provided by the infrared dynamics of QCD,
 yields two boundary conditions which constrain 
 the corresponding eigenvalues $\lambda_i$ to take discrete values. 
 This results in an isolated pole for the QCD Pomeron as opposed to 
 the cut obtained for fixed $\alpha_s$.
 
 \noindent
 The necessity of introducing the running of the coupling is at
 first sight surprising since the BFKL kernel is infrared finite.
 This means that if the transverse momentum of the $t$-channel gluons 
 at one end of the exchanged ladder is fixed by the impact factor
 describing its coupling to the scattering hadron,
 then  the typical transverse momenta of the gluons in the subsequent 
 sections of the ladder will be of the same order of magnitude. 
 However, as explained in ref.\cite{lip86}, the diffusion in 
 transverse momentum along the ladder means that a wider 
 range of  transverse momenta contributes to the BFKL amplitude,
 so that the amplitude is indeed sensitive to the argument of the running
 coupling at each rung of the ladder.

 \noindent
 On the other hand, the precise mechanism used to encode the
 running of the coupling has little effect on the behaviour 
 of the eigenfunctions at the critical value of transverse momentum,
 where the function changes from oscillation to exponential decay. 
 In other words, it makes little difference  whether the argument 
 of the coupling at a particular rung is set equal 
 to the transverse momentum above or below that rung. 
 In terms of the integral equation this means that one has the freedom
 to set the argument of the coupling equal to the external
 (unintegrated) value, $\vec{k}$, rather than the integrated one 
 $\vec{k'}$, and write eq.(\ref{eq1})  as
 \begin{equation}
 \hspace{-0.95cm}
 \alpha_s(\vec{k}^2) 
 \int \d^2\vec{k'} \, {\mathcal K}(\vec{k},\vec{k'}) \,
 \phi_i(\vec{k'})  = \lambda_i \, \phi_i(\vec{k}).
 \label{eq2} 
 \end{equation} 
 The reason for this is that although far enough along the ladder 
 we expect to have to consider transverse momenta whose magnitude 
 differs substantially from the value set by the impact factors, 
 the transverse momenta of two adjacent sections of the ladder are
 indeed of the same order. 
 As far as (renormalisation group improved) perturbation 
 theory is concerned, therefore, a discussion about which transverse 
 momentum should be used in the running of the coupling will only
 affect  the solutions to the BFKL equation at the 
 subleading logarithm level.

 \noindent
 However, if one wishes to consider in more detail the 
 infrared contributions to BFKL
 and run the coupling down to the region of the Landau pole,
 where small changes in  the argument lead to substantial changes 
 in the value of the coupling, one no longer has the above freedom 
 and a study of the precise prescription for running the coupling 
 becomes pertinent. 
 Such a precise prescription is necessary for an analysis 
 of the renormalon structure of the BFKL amplitude aimed at
 identifying  non-perturbative power corrections to the
 amplitude which are likely to play a crucial role in the
 reconciliation of the perturbative QCD Pomeron with the phenomenological
 ``soft'' Pomeron.

 \noindent
 In a recent publication \cite{levin} Levin has conjectured that
 the correct prescription for the coupling is
 \begin{eqnarray}
 \hspace{-0.95cm} \frac{\alpha_s (\vec{k'}^2) \,
 \alpha_s((\vec{k}-\vec{k'})^2)}{\alpha_s(\vec{k}^2)}. \nonumber  
 \end{eqnarray}
 His analysis, however, was based on the idea of considering the running of 
 the coupling at one rung of the ladder only, whilst keeping the 
 coupling fixed throughout the rest of the ladder. 
 A  systematic treatment requires a prescription which can be extended
 to the entire ladder. 
 In this letter we propose  just such a systematic treatment
 in order to arrive at a well-defined algorithm for the introduction of
 the running of the coupling.
 
 \noindent
 Our starting point is not the gluon ladder approach of BFKL 
 \cite{bfkl} but rather the dipole expansion of Mueller 
 \cite{mueller} and  Nikolaev and  Zakharov \cite{nz}.
 The difference between the two approaches lies in the description of
 the {\it same} high energy (or small $x$) physical process.
 In the BFKL approach the process occurs via the exchange of a ladder
 of reggeized gluons in the $t$-channel. The leading logarithmic
 corrections in the energy are generated by the increasing  
 number of rungs along the gluon ladder. 
 The exchanged gluons, though, cannot be interpreted in a
 probabilistic fashion as emanating from a specific hadron involved in
 the process. 
 The gluon ladder does not have a straightforward parton model
 interpretation with an identifiable  hard scattering section.
 In the dipole approach the process occurs via the sequential
 emission of colour dipoles in the $s$-channel by the participating
 hadrons.
 The leading logarithmic corrections in the energy are generated
 {\it radiatively} by the increasing number of emitted dipoles.
 This dipole cascade can be interpreted in the appropriate gauge
 (light cone gauge) as a component of the hadronic wavefunction.
 The hard subprocess is encoded in the dipole cross section
 that describes the scattering of two dipoles 
 for onium-onium elastic scattering, or the absorption by the
 nucleon target of a dipole emitted by the virtual photon for
 small $x$ D.I.S. 
 In impact parameter space, where transverse separations 
 $\vec{x}$ and $\vec{x \,}'$  replace the transverse momenta
 $\vec{k}$ and $\vec{k}'$, the factorization of the 
 dipole cross section from the squared wave function for 
 dipole emission takes a particularly simple form. 

 \noindent
 The equivalence of the two approaches for fixed $\alpha_s$
 and for inclusive observables such as  onium-onium total cross
 section and $F_2$ in small $x$ D.I.S. has been shown in refs. 
 \cite{mueller2} and \cite{nzz} respectively.
 For a systematic inclusion of the running coupling we find
 the dipole approach more suitable exactly because of its
 radiative nature.  
 Our method is the following.
 The emission of a daughter dipole by a parent dipole in
 momentum space occurs via the radiation of a soft gluon
 with transverse momentum $\vec{k_1}$ as shown in fig.1.
 We take the argument of the running coupling in this case
 to be equal to $\vec{k_1}^2$, in accordance with the treatment
 of soft gluon radiation in the case of timelike cascades
 \cite{basics}.  
 We note that the relevant kinematic region for the momentum of 
 the soft gluon is the Glauber region where its virtuality 
 is determined by its transverse momentum, i.e.
 \mbox{ $ k_1^2 \approx - \vec{k_1}^2$}.
 To translate this running into impact parameter space representation
 we require that the Borel transformed probability for emission 
 of a soft gluon with impact parameter $\vec{y_1}$ 
 from the parent dipole with size $\vec{x}$ 
 should be equal to the expression obtained by first Borel and Fourier
 transforming the two graphs of fig.1 and then squaring their sum.
 Since this procedure correctly reproduces the singularity structure in
 the Borel plane, it provides a prescription for the coupling that
 allows us to run down towards the Landau pole.
 It turns out that the argument of the running of the coupling 
 cannot be expressed simply as a function of $\vec{x}$ and $\vec{y_1}$
 only,  but the expression we obtain nevertheless  has a convenient
 integral representation. 

 \begin{figure}
 \begin{center}
 \leavevmode
 \hbox{\epsfxsize=4.4 in 
 \epsfbox{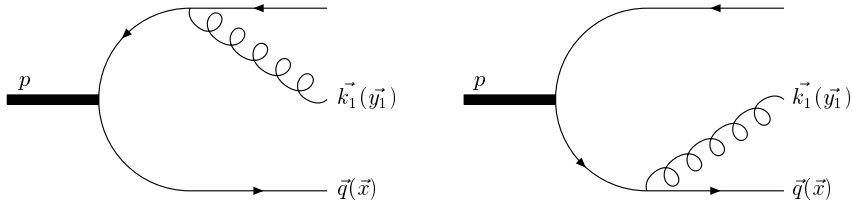}}
 \end{center}
 \caption{Graphs for the emission of a soft gluon with
 transverse momentum $\vec{k_1}$ (impact parameter $\vec{y_1}$)
 from a dipole with momentum $p$, in which the quark has transverse momentum
 $\vec{q}$ (impact parameter $\vec{x}$).} \label{fig1}
 \end{figure}
 
 \noindent
 To this end we consider the
 emission of a soft gluon with momentum $k_1^\mu$, colour $a$, 
 polarization $\epsilon$ and longitudinal momentum fraction $z_1$,
 from a fast moving dipole with total momentum  $p^\mu$,
 in which the quark and antiquark have momenta $q^\mu$ and
 $(p-k_1-q)^\mu$ and $m_q$ is the quark mass,
 \begin{eqnarray}
 \hspace{-0.95cm}
 q^\mu = (p \, z , \frac{m_q^2+\vec{q^2}}{p \, z}, \vec{q}). \nonumber
 \end{eqnarray}
 Leading logarithmic contributions in the energy are obtained 
 in the kinematic region of strongly ordered rapidities
 \mbox{ $z_1 \ll z $.}
 The amplitude for such a process with running coupling as described
 above is given by \cite{mueller}
 \begin{equation}
 \hspace{-0.95cm}
 \psi^{(1)a}(\vec{q},\vec{k_1},z,z_1)  = -2 g(\vec{k_1}^2) T^a \left[
 \psi^{(0)}(\vec{k_1},z)-\psi^{(0)}((\vec{k_1}+\vec{q} \, ),z) \right]
 \frac{ \vec{k_1} \cdot \epsilon} {\vec{k_1}^2},  
 \label{eq3} 
 \end{equation} 
 where $\psi^{(0)}$ is the wavefunction for the dipole without gluon 
 emission and $T^a$ is a colour generator in the adjoint
 representation.
 We now introduce the Borel transformed amplitude
 $\tilde{\psi}^{(1)a}$ defined by 
 \begin{equation}
 \hspace{-0.95cm}
 \psi^{(1)a}(\vec{q}, \vec{k_1}, z, z_1) =
 \int_0^\infty \d b \, \tilde{\psi}^{(1) a}( \vec{q}, \vec{k_1}, z,
 z_1, b) \, e^{-b/ \alpha_s(M^2)}.
 \label{boreldef}
 \end{equation}
 Note that the Borel dual variables are $b$ and $\alpha_s(M^2)$,
 where $M^2$ is an overall external momentum scale to be determined
 once the dipoles are considered in the context a specific process.
 For D.I.S. at small $x$, for instance, $M^2$ is the spacelike
 virtuality $Q^2$ of the
 photon $\gamma^\star$, which fluctuates into the dipole  cascade
 before the interaction with the nucleon target.
 The Borel transform of the wavefunction in eq.(\ref{eq3}) is
 \begin{equation}
 \hspace{-0.95cm} 
 \tilde{\psi}^{(1)a}(\vec{q}, \vec{k_1}, z, z_1, b) 
  = -2 \tilde{g}(\vec{k_1}^2, b) T^a \left[
 \psi^{(0)}(\vec{k_1},z)-\psi^{(0)}((\vec{k_1}+\vec{q}),z) \right]
 \frac{ \vec{k_1} \cdot \epsilon} {\vec{k_1}^2},
 \label{borelpsi}
 \end{equation}
 where
 \begin{equation}
 \hspace{-0.95cm}
 \tilde{g}(\vec{k_1}^2, b) = \frac{2}{\sqrt{b}} 
 \left( \frac{\vec{k_1}^2}{M^2}  \right)^{-b \beta_0}, 
 \label{borelg} 
 \end{equation}
 and $\beta_0 = (1/4\pi)((11/3)N_c -(2/3)N_f)$.
 The Borel transformed $\tilde{g}$ is determined by the requirement
 that the ``square'' of $ \tilde{g}(\vec{k_1}^2,b)$ in the sense of
 Borel convolution with itself
 results in the Borel transform  of $4 \pi \alpha_s(\vec{k_1}^2)$, i.e.  
 \begin{eqnarray}
 \hspace{-0.95cm} 
 4 \pi \tilde{\alpha_s}(\vec{k_1}^2, b)= \left(
 \frac{\vec{k_1}^2}{M^2}  \right)^{-b \beta_0}  \hspace{-0.2cm} & = & 
 \tilde{g}(\vec{k_1}^2, b) \otimes \tilde{g}(\vec{k_1}^2,b) \nonumber \\
 & \equiv &  \int_0^b \d b' \, \tilde{g}(\vec{k_1}^2, b')
 \tilde{g}(\vec{k_1}^2,(b-b')) .
 \label{conv} 
 \end{eqnarray}
 Taking the Fourier transform of eq.(\ref{borelpsi}) 
 we obtain the wavefunction as a function of impact parameters 
 $\vec{x}, \, \vec{y_1}$ conjugate to $\vec{q}, \, \vec{k_1}$ respectively
 \begin{eqnarray}
 \hspace{-0.95cm}
 \tilde{\psi}^{(1)a}(\vec{x},\vec{y_1},z,z_1,b) &  =& 
 - \frac{1}{\sqrt{b}} \frac{2 {\mathrm i}}{\pi} T^a
 \psi^{(0)}(\vec{x},z) \int_0^\infty \d\tau \, {\mathrm J}_1(\tau)
 \nonumber \\
 & &  \hspace*{-1.95cm}
 \times \left[ \left( \frac{M^2 \vec{y_1}^2}{ \tau^2} \right)^{b
 \beta_0} \frac{\epsilon \cdot \vec{y_1}}{\vec{y_1}^2} + \left(
 \frac{M^2 (\vec{x}-\vec{y_1})^2}{ \tau^2} \right)^{b  \beta_0 } 
 \frac{\epsilon \cdot (\vec{x}-\vec{y_1})}{(\vec{x}-\vec{y_1})^2} \right] ,
 \label{fourierpsi} 
 \end{eqnarray}
 where ${\mathrm J}_1$ is the first order Bessel function.
 The Borel transform, $\tilde{\Phi}(\vec{x},z,b)$,
 of the total probability for the emission of a gluon
 from a dipole separated by impact parameter $\vec{x}$ in which
 one of the fermions carries a fraction of longitudinal momentum $z$, is
 obtained by  taking the ``square'' of the amplitude 
 given by eq.(\ref{fourierpsi}). By ``square'' we  mean 
 convolution of $\tilde{\psi}^{(1)a}$ in $b$ with itself (as defined in
 eq.(\ref{conv})) and integration over the phase space of the outgoing gluon. 
 This gives
 \begin{eqnarray}
 \hspace{-0.95cm}
 \tilde{\Phi}^{(1)}(\vec{x},z,b) & =&   \frac{C_F}{\pi^3} \int^z  
 \frac{\d z_1}{z_1}  
 \int \d^2 \vec{y_1}   
 \int_0^\infty \d \tau \d \tau' {\mathrm J}_1(\tau) { \mathrm J}_1(\tau')
 \int_0^1 \frac{\d \omega}{\omega^{\frac{1}{2}}
 (1-\omega)^{\frac{1}{2}}}   
 \nonumber \\ 
 & \times \,&  \left(  \frac{1}{\tau^2} \right)^{\omega b \beta_0}  
 \left(  \frac{1}{\tau^{\prime 2}} \right)^{(1-\omega)b \beta_0}
 \frac{(M^2)^{b \beta_0}}{\vec{y_1}^2  (\vec{x}-\vec{y_1})^2}
 \nonumber \\ 
 & \times \, &  \left[ ( \vec{y_1}^2)^{b \beta_0}
 (\vec{x}-\vec{y_1})^2  + ((\vec{x}-\vec{y_1})^2 )^{b  \beta_0}
 \vec{y_1}^2 \right. \nonumber \\ 
 & & \, \left. +2  \vec{y_1} \cdot  (\vec{x}-\vec{y_1}) (\vec{y_1}
 ^2)^{\omega b \beta_0}  ((\vec{x}-\vec{y_1})^2
 )^{(1-\omega)b  \beta_0}  \right]   \Phi^{(0)}(\vec{x},z) , 
 \label{borelphi} 
 \end{eqnarray} 
 where $\Phi^{(0)}$ is the probability for the dipole without gluon emission.
 The integration over the variables $\tau, \, \tau'$ can be readily
 performed. 
 However if we choose {\it not} to do this then each term in the
 integrand can be identified as the Borel transform of a running coupling
 with a suitable argument. 
 Thus we define
 \begin{eqnarray}
 \hspace{-0.95cm}
 \vec{r_1}^2(\omega) &=& (\vec{y_1}^2)^{\omega } ( (\vec{x}-\vec{y_1})^2
 )^{(1-\omega)} \nonumber 
 \end{eqnarray}
 \begin{eqnarray}
 \hspace{-0.95cm}
 \lambda^2(\tau,\tau' \omega) &=&
  (\tau^2)^{\omega } ( \tau^{\prime 2} )^{(1-\omega)} , \nonumber
 \end{eqnarray}
 and take the inverse Borel transform of eq.(\ref{borelphi}) to obtain
 the following expression for the probability for gluon emission 
 with running coupling
 \begin{eqnarray}
 \hspace{-0.95cm}
 \Phi^{(1)}(\vec{x},z) &=&  \frac{C_F}{ \pi^3} \ln(z)  
 \int \d^2\vec{y_1} 
 \int_0^\infty \d \tau \d \tau' { \mathrm J}_1(\tau) { \mathrm J}_1(\tau')
 \int_0^1 \frac{ \d \omega}{\omega^{\frac{1}{2}} (1-\omega)^{\frac{1}{2}}}
 \nonumber \\  
 & & \hspace*{-2.1cm} \times \, 
 \frac{1}{\vec{y_1}^2  (\vec{x}-\vec{y_1})^2} \left[
 \alpha_s(\lambda^2/\vec{r_1}^2) \vec{x\,}^2  +\left( \alpha_s(\lambda^2/\vec{
 y_1}^2)-\alpha_s(\lambda^2/\vec{r_1}^2) \right) 
 (\vec{x}-\vec{y_1})^2
 \right. 
 \nonumber \\   
 & & \hspace*{0.9cm} +  \left. \left(
 \alpha_s(\lambda^2/(\vec{x}-\vec{y_1})^2)-\alpha_s(\lambda^2/\vec{ 
 r_1}^2) \right)  \vec{y_1}^2   \right] \Phi^{(0)}(\vec{
 x},z). \label{phi}
 \end{eqnarray}
 In the limit where the coupling is fixed the integrations over
 $\tau$, $\tau'$ , $\omega$ give a factor of $\pi$ and the first term
 on the R.H.S. of eq.(\ref{phi}) coincides with the expression
 obtained in the usual treatment of dipole radiation (see, for example 
 eq.(10) of ref.\cite{mueller}) whereas the other two  terms vanish.
 For running couplings we are unambiguously led to the prescription of
 eq.(\ref{phi}), namely that the kernel has three terms
 each of which has a different running coupling and for each of
 these terms the argument of the running coupling is {\it not}
 expressed simply in terms of the impact parameters $\vec{x}$
 and $\vec{y}$, but also in terms of three further parameters,
 $\tau, \tau', \omega$, the emission probability
 being a weighted integral over these parameters.
 It is worth noting that the weight function peaks when
 $\lambda(\tau,\tau', \omega)$ is equal to unity, so that although
 one has to integrate over all possible arguments for the running coupling,
 the integral is highly peaked in the region where the running
 depends only on  $\vec{y_1}^2$ or $(\vec{x}-\vec{y_1})^2$ or
 some mean of the two. We note that had we taken the Fourier transform
 of the wavefunction $\psi^{(1)a}$ in eq.(\ref{eq3}) and squared it to
 construct $\Phi^{(1)}$ as in eq.(\ref{phi}), we would have ended up
 with a product of the form $g(\tau/\vec{y \, }^2) g(\tau'/ (\vec{x} -
 \vec{y}\,)^2)$. It is the use of the Borel representation that enables
 us to 
 express this product in terms of running $\alpha_s$ of a single
 argument.  This then is a consistent and systematic way to introduce 
 the running of the coupling, valid to all orders in perturbation theory.

 \noindent
 So far we have dealt with the wavefunction squared $\Phi$ without
 considering its coupling to some hard scattering subprocess.
 Such a coupling was considered in ref.\cite{mueller2} where,
 through the use of time ordered perturbation theory, 
 it was shown that final state interactions cancel and are therefore 
 not included in the dipole kernel.
 Consequently, final state interactions will not affect the method we suggest
 for the running of the coupling. 

 \noindent
 Returning to the Borel transformed  $\tilde{\Phi}^{(1)}$, in
 eq.(\ref{borelphi}), the integrations over $\vec{y_1}$, 
 $\tau$, $\tau'$, $\omega$ may be performed to yield 
 \begin{eqnarray}
 \hspace{-0.95cm}
 \tilde{\Phi}^{(1)}(\vec{x},z,b) = -\frac{2 C_F}{\pi} \ln (z)
 \frac{\Gamma(-b \beta_0)}{\Gamma(1+b \beta_0)} \left( \frac{M^2 \vec{
 x}^2 }{4} \right)^{b\beta_0} \Phi^{(0)}(\vec{x},z) , \label{phi_one}
 \end{eqnarray}
 which has the usual infrared renormalon poles for positive integer values
 of $b \beta_0$ \cite{renorm}.  
 The poles at $b=0$ and for negative integer
 values of $b \beta_0$ are  ultraviolet renormalon poles and are
 eliminated if a short distance cutoff is introduced for the
 integration over $\vec{y_1}$. Unitarity tells us that the one gluon
 exchange virtual contribution is $-\tilde{\Phi}^{(1)}(\vec{x},z, b)$.

 \noindent
 If we  square the amplitude for single
 gluon emission calculated in momentum space, take the Borel transform of
 the running coupling and then  take a Fourier transform 
 in the transverse momentum $\vec{q}$ of the fermion {\it only} 
 we get upon integrating over all phase-space for the outgoing gluon
 \begin{equation}
 \hspace{-0.95cm}
 \tilde{\Phi}^{(1)}( \vec{x}, z, b) =
 \frac{C_F}{\pi^2} \ln(z) 
 \int \frac {\d^2\vec{k_1}}{\vec{k_1}^2}  \,
 \left( \frac{M^2}{\vec{k_1}^2} \right)^{b \beta_0} 
 \left[ 2-e^{i \vec{k_1} \cdot \vec{x}}-e^{-i \vec{k_1} \cdot \vec{x}} \right] 
     \Phi^{(0)}(\vec{x},z) \, , 
 \label{borelmom}
 \end{equation}
 which also yields eq.(\ref{phi_one}) when the integral over $\vec{k_1}$ is 
 performed.  This is not surprising but it serves as a check and 
 confirms that one is at liberty to commute Fourier and Borel 
 transforms.

 \noindent
 In exact analogy with the treatment of ref.\cite{mueller} the process can
 be iterated to account for the emission of any number of soft gluons.
 To this end we define the generating functional
 $\tilde{Z} \left( \vec{x}, \, \vec{0} ,z,j(\vec{y\,}',z' ),b \right)$ 
 to be the quantity whose   $n^{th}$ functional derivative with
 respect to the source, $j(\vec{y\,}',z')$,
 generates the Borel transform of the probability   for the emission of
 $n$ gluons.  Due to the exponentiation of soft gluon radiation,
 virtual corrections can be accounted for by the factor $\exp \left( -
 \tilde{\Phi}^{(1)} (\vec{x} ,z ,b ) \right)$ \cite{basics}.  Then the
 generating functional $\tilde{Z}$ obeys the Bethe-Salpeter equation 
 \begin{eqnarray}
 \hspace{-0.95cm}
 \tilde{Z} \left(\vec{x}, \, \vec{0},z,j,b \right) &=& \exp \left(
 \frac{2 C_F}{\pi} \ln (z) \frac{\Gamma(-b \beta_0)}{\Gamma(1+b
 \beta_0)}  \left( \frac{M^2\vec{x\,}^2}{4} \right)^{b\beta_0} \right)
 \nonumber \\
 & & \hspace*{-3.3cm}
 + \frac{ C_F}{\pi^3} \int^z \frac{\d z'}{z'} \exp \left( \frac{2
 C_F}{\pi} \ln (z/z') \frac{\Gamma(-b \beta_0)}{\Gamma(1+b \beta_0)}
 \left( \frac{M^2\vec{x\,}^2}{4} \right)^{b \beta_0 } \right)
 \nonumber \\ 
 & & \hspace{-2.9cm}
 \otimes  \int \d^2 \vec{y \,}' \int_0^\infty \d \tau \d \tau'
 { \mathrm J}_1(\tau) {\mathrm J}_1(\tau') \int_0^1
 \frac{\d \omega}{\omega^{\frac{1}{2}} (1-\omega)^{\frac{1}{2}}} \left(
 \frac{1}{\tau^2} \right)^{\omega b \beta_0} \left(
 \frac{1}{\tau^{\prime 2}} \right)^{(1-\omega)b \beta_0}  \nonumber \\ 
 & & \hspace{-2.6cm}
 \frac{M^{2b \beta_0}}{{\vec{y \,}' \,}^2  {(\vec{x}-\vec{y \,}')^2}}
 \left[ ( \vec{y \,}'^2)^{b \beta_0} {(\vec{x}-\vec{y \,}') }^2
 +({(\vec{x}-\vec{y \,}')}^2 )^{b \beta_0} \vec{y \,}'^2   \right.
 \nonumber \\ 
 & & \hspace{-0.2cm}
 \left. +2 \vec{y \,}' \cdot  (\vec{x}-\vec{y \,}') (\vec{y
 \,}'^2)^{\omega b  \beta_0 }(\vec{x}-\vec{y \,}')^2 )^{(1-\omega)b
 \beta_0}  \right] \nonumber  \\ 
 & & \hspace{-0.2cm} 
 \otimes \, \tilde{Z} \left( \vec{y \,}', \, \vec{x },z',j,b\right)
 \otimes \tilde{Z} \left( \vec{y \,}', \, \vec{0} ,z',j,b \right) \,
 j(\vec{y \,}',z'), 
 \label{genfun}
 \end{eqnarray}
 where the exponentials of functions of $b$ are to be understood in the
 sense of their Taylor expansion with products replaced by convolutions,
 i.e.
 \begin{eqnarray}
 \hspace{-0.95cm}
 \exp \left( f(b) \right) = \delta(b)+f(b)+\frac{1}{2 !} f(b)
  \otimes f(b) + \cdots \nonumber 
 \end{eqnarray}
 \noindent
 The $n^{th}$ functional derivative of eq.(\ref{genfun}) with respect to 
 the source $j$ is equivalent to the $n^{th}$ iteration 
 of the integral equation
 \begin{equation}
 \hspace{-0.95cm}
 z \frac{\partial}{\partial z} \tilde{\Phi}(x,z,b) = z
 \frac{\partial}{\partial z} \Phi^{(0)} (x,z) \delta (b)+  \int_0^\infty \d x' 
 \tilde{{\mathcal K}}(x,x',b) \otimes \tilde{\Phi}(x',z,b),
 \label{eqnew}
 \end{equation}
 where $\tilde{{\mathcal K}}$ is the Borel transformed dipole kernel 
 and a change of variables has been affected  from
 the two dimensional vector $\vec{y \,}'$ to its modulus $y'$
 and the modulus, $x'$,  of  $\vec{x \,}'=(\vec{x}-\vec{y \,}')$.
 The Jacobian for this change of variables can be written as \cite{mueller}
 \begin{equation}
 \hspace{-0.95cm}
 \d^2 \vec{y \,}' = 2 \pi \, y' \, x' \d y' \d x' 
 \int_0^\infty \kappa \d \kappa 
 {\mathrm J}_0(\kappa x) \,  {\mathrm J}_0(\kappa x' ) \,
 {\mathrm J}_0(\kappa y'), 
 \label{jacobian}  
 \end{equation}
 where ${\mathrm J}_0$ are Bessel functions of zeroth order. Extracting
 $\tilde{\mathcal{K}}$ from eq.(\ref{genfun}) and integrating over
 $y'$ and $\kappa$, we find
 \begin{eqnarray}
 \hspace{-0.95cm}
 \tilde{{\mathcal K}}(x,x',b) &=& - \frac{C_A}{\pi} \left( \frac{M^2
 x^2}{4} \right)^{b \beta_0} \frac{\Gamma(-b \beta_0 )}{\Gamma(1+ b
 \beta_0)} \delta (x - x') \nonumber \\ 
 & & \hspace{-2.6cm}
 + \frac{C_A}{\pi^2} \frac{1}{x'}
 \left( \frac{M^2 x'^2}{4} \right)^{b \beta_0} \int_0^1
 \frac{\d\omega}{\omega^{\frac{1}{2}} (1-\omega)^{\frac{1}{2}}}
 \frac{\Gamma(1-(1-\omega)b \beta_0) \, \, \Gamma(1-\omega b \beta_0
 )}{\Gamma(1+ (1-\omega) b \beta_0 ) \Gamma(1 + \omega b \beta_0 )}
 \nonumber \\
 & & \hspace{-2.3cm}
 \times \, \left\{ \left( \frac{x_>^2}{x'^2} \right)^{b \beta_0 - 1}
 {\mathrm F} \left( 1- b \beta_0 , 1- b \beta_0 ; 1 ;
 \frac{x_<^2}{x_>^2} \right) \right. \nonumber \\ 
 & & \hspace{-1.4cm}
 \left. +  \left( \frac{x^2 - x'^2}{x_>^2}
 \right) \left( \frac{x_>^2}{x'^2}  \right)^{\omega b \beta_0}
 {\mathrm F} \left( 1 - \omega b \beta_0 , 1 - \omega b \beta_0 ; 1 ;
 \frac{x_<^2}{x_>^2} \right) \right. \nonumber \\
 & & \hspace{-1.4cm}
 \left. - \left(\frac{x_>^2}{x'^2} \right)^{\omega  b \beta_0} 
 {\mathrm F} \left( - \omega b \beta_0 , - \omega b \beta_0 ; 1 ; 
 \frac{x_<^2}{x_>^2} \right) \right\} \, ,
 \label{borelK}
 \end{eqnarray}
 where $x_< = {\mathrm min} (x,x')$, $x_> = {\mathrm max} (x,x')$ and
 ${\mathrm F}$ is the Gauss hypergeometric function. We
 have replaced $C_F$ by $\frac{1}{2}C_A$, which is the colour factor 
 occurring in the BFKL equation and agrees with $C_F$ in the leading 
 $1/N_c$ approximation. The first term of
 eq.(\ref{borelK}) describes the virtual corrections and it is
 proportional to $\delta (x - x')$ due to the completeness of the
 Bessel functions.

 \noindent
 The eigenfunctions, $\tilde{\phi}(x,b)$ of this kernel are also
 functions of $b$ so that the action of the kernel is also to be 
 understood in the sense of a convolution, i.e. the eigenvalue equation is
 \begin{equation}
 \hspace{-0.95cm}
  \int_0^\infty \d x' \int_0^b \d b' 
  \tilde{{\mathcal K}}(x,x',b')
  \tilde{\phi}_i(x',(b-b'))  
  = \lambda_i(b) \, \tilde{\phi}_i(x,b) .
  \label{eigen}  
 \end{equation}

 \noindent
 The eigenvalues and eigenfunctions of this kernel must be
 found in order to investigate the analyticity structure of the Borel
 transform of the BFKL amplitude. 
 Clearly eq.(\ref{eigen}) does not lend itself 
 to a simple analytic solution. The eigenfunctions of the kernel 
 will have to be examined by numerical methods
 in order to locate their singularities. 
 The location of these singularities will give us information about 
 the non-perturbative contributions to the BFKL amplitude. 
 We shall report on the results of such an analysis 
 in a forthcoming publication.

 \noindent
 {\it Acknowledgement:} \,
 The authors would like to thank A. H. Mueller for insightful
 discussions on the dipole approach. Two of us (KDA, MGS) would like
 to thank PPARC for their financial support.


\newpage

 \begin{thebibliography} {99}

 \bibitem{lip86} L. N. Lipatov, Sov. Phys. JETP 63 (1986) 904

 \bibitem{bfkl} E. A. Kuraev, L. N. Lipatov and V. S. Fadin,
               Sov. Phys. JETP 45 (1977) 199 \\
               Ya. Ya. Balitsky and L. N. Lipatov,
               Sov. J. Nucl. Phys. 28 (1978) 822 

 \bibitem{levin} E. Levin, Nucl. Phys. B453 (1995) 303

 \bibitem{mueller} A. H. Mueller, Nucl. Phys. B415 (1994) 373


 \bibitem{nz} N. N. Nikolaev and B. G. Zakharov, Z. Phys. C64 (1994) 631 


 \bibitem{mueller2} Z. Chen and A. H. Mueller, Nucl. Phys. B451 (1995) 579

 \bibitem{nzz} N. N. Nikolaev, B. G. Zakharov and V. R. Zoller, Sov.
               Phys. JETP Lett. 59 (1994) 6  

 \bibitem{basics} Yu. L. Dokshitzer, V. A. Khoze, A. H. Mueller
                 and S. I. Troyan, `` Basics of Perturbative QCD'',
                 ed. Fronti\`{e}res, Gif-sur-Yvette, 1991
  

 \bibitem{renorm}
       A. H. Mueller, Nucl. Phys. B250 (1985) 327 \\
       G. P. Korchemsky and G. Sterman, Nucl. Phys. B437 (1995) 415 \\ 
       M. Beneke and V. M. Braun, Nucl. Phys. B454 (1995) 253

 \end{thebibliography}
 \end{document}